# Tachyonic equations to reduce the divergent integral of QED

Z.Y. Wang

*Research Center, No.580 Songhuang Road,
Shanghai 201703, P.R. China*

**Abstract:** The momentum interval $0\sim\infty$ only represents the velocity lower than that of light, i.e. $0<V<c$ of special relativity. If the total integral including both $0<V<c$ and $c<V<\infty$ is considered, divergence will be reduced. In this paper, we firstly use the Cherenkov radiation in media as analogue to confirm the formulation (7)~(10) of a tachyon in vacuum. Then these equation are applied to calculate the Lamb shift and the cut-off frequency is unnecessary

# 1. Relativistic formula to derive Cherenkov effect in media

The angle $\cos\theta = \dfrac{c/n}{V}$ given by electrodynamics can also be deduced from relativistic mechanics. The momentum conservation is

$$\mathbf{p}_1 - \hbar\mathbf{k} = \mathbf{p}_2$$

where $\hbar\mathbf{k}$ is the momentum of the emitted photon, $\mathbf{p}_1$ and $\mathbf{p}_2$ represents the initial and final momentum of the charged particle. Thus,

$$p_1^2 - 2p_1\hbar k \cos\theta + \hbar^2 k^2 = p_2^2$$

$$\cos\theta = \frac{p_1^2 + \hbar^2 k^2 - p_2^2}{2p_1\hbar k} \quad (1)$$

As to energy conservation,

$$E_1 - \hbar\omega = E_2 \quad (2)$$

If the energy is $E_1 = \sqrt{p_1^2 c^2 + m_0^2 c^4}$ and $E_2 = \sqrt{p_2^2 c^2 + m_0^2 c^4}$ of special relativity in Equ.(2),

$$p_1^2 c^2 + m_0^2 c^4 - 2\hbar\omega E_1 + \hbar^2\omega^2 = p_2^2 c^2 + m_0^2 c^4$$

$$p_1^2 - 2\hbar\omega E_1 / c^2 + (\hbar\omega / c)^2 = p_2^2 \quad (3)$$

Substituting (3) into (1),

$$\cos\theta = \frac{\hbar^2 k^2 - (\hbar\omega/c)^2 + 2\hbar\omega E_1/c^2}{2p_1\hbar k} = \frac{\hbar^2 k^2 - (\hbar\omega/c)^2}{2p_1\hbar k} + \frac{\omega}{k}\frac{E_1/c^2}{p_1} \quad (4)$$

Since $\omega/k = c/n$ to light in the medium and $\dfrac{E_1/c^2}{p_1} = \dfrac{1}{V_1}$, it is

$$\cos\theta = \frac{\hbar\omega \ (n^2 - 1)}{2p_1 cn} + \frac{c/n}{V_1} \quad (5)$$

which is inconsistent with the known relation $\cos\theta = \dfrac{c/n}{V_1}$ of electrodynamics. A solution to the problem was introduced[1]. According to the postulate, the initial and final energy of the particle should now be $E_1 = \sqrt{p_1^2 u^2 + m_0^2 u^4}$ and $E_2 = \sqrt{p_2^2 u^2 + m_0^2 u^4}$ ($u = c/n$ is the light speed in media). Hence,

$$p_1^2 u^2 + m_0^2 u^4 - 2\hbar\omega E_1 + \hbar^2\omega^2 = p_2^2 u^2 + m_0^2 u^4 \quad (2')$$

$$p_1^2 - 2\hbar\omega E_1 / u^2 + (\hbar\omega / u)^2 = p_2^2 \quad (3')$$

Make a comparison with (1) and (3'),

$$\cos\theta = \frac{\hbar^2 k^2 - (\hbar\omega/u)^2}{2p_1\hbar k} + \frac{\omega}{k}\frac{E_1/u^2}{p_1} \quad (4')$$

Owing to $\omega/k = c/n = u$ and $\dfrac{E_1/u^2}{p_1} = \dfrac{1}{V_1}$, the angle is

$$\cos\theta = \frac{\omega}{k}\frac{E_1/u^2}{p_1} = \frac{u}{V_1} = \frac{c/n}{V_1} \quad (5')$$

or $\quad \sin\varphi = \dfrac{c/n}{V_1} \quad (\varphi = \dfrac{\pi}{2} - \theta) \quad (6)$

Actually, in aerodynamics there has a similar phenomenon to the Cherenkov effect if the object is faster than the sonic speed $c_s$ where $\varphi = \sin^{-1}\dfrac{c_s}{V_1}$ is the so-called Mach angle. It can also be interpreted by these equations provided the photon is replaced by a phonon and $c/n \longrightarrow c_s$.[2]

## 2. Superluminal equations to derive

The defect of the above method is that the Cherenkov effect occurs when $V > u$ but the equations based on relativistic mechanics are invalid in this case for the implicit factor $\sqrt{1 - V^2/u^2}$ is imaginary. We assume the formula of momentum and energy on condition that $V > u$ to be

$$E = \frac{m_0 u^2}{\sqrt{V^2/u^2 - 1}}$$

$$p = \frac{m_0 V}{\sqrt{V^2/u^2 - 1}}$$

$$E^2 = p^2 u^2 - m_0^2 u^4$$

$$\frac{E}{p} = \frac{u^2}{V}$$

Therefore, the initial and final energy in (2) should be $E_1 = \sqrt{p_1^2 u^2 - m_0^2 u^4}$ and $E_2 = \sqrt{p_2^2 u^2 - m_0^2 u^4}$,

$$p_1^2 u^2 - m_0^2 u^4 - 2\hbar\omega E_1 + \hbar^2\omega^2 = p_2^2 u^2 - m_0^2 u^4 \quad (2'')$$

$$p_1^2 - 2\hbar\omega E_1/u^2 + (\hbar\omega/u)^2 = p_2^2 \quad (3')$$

Then the procedure from (4') to (6) can be repeated and the result is still correct. Obviously, it is also tenable to supersonic phenomena such as the Mach angle.

In vacuum, the light speed $u = c/n$ of these equations must be replaced by $c$ as

$$E = \frac{m_0 c^2}{\sqrt{V^2/c^2 - 1}} \tag{7}$$

$$p = \frac{m_0 V}{\sqrt{V^2/c^2 - 1}} \tag{8}$$

$$E^2 = p^2 c^2 - m_0^2 c^4 \tag{9}$$

$$\frac{E}{p} = \frac{c^2}{V} \tag{10}$$

$$\cos\theta = \frac{c}{V}$$

The consequential spacetime transformations are

$$x' = \frac{x - Vt}{\sqrt{V^2/c^2 - 1}}$$

$$y' = y$$

$$z' = z$$

$$t' = \frac{t - xV/c^2}{\sqrt{V^2/c^2 - 1}}$$

and the addition rule of velocities is

$$\frac{dx'}{dt'} = \frac{dx/dt - V}{1 - \frac{V}{c^2}\frac{dx}{dt}}$$

Thereby, $c' = \dfrac{c - V}{1 - \dfrac{Vc}{c^2}} = c$ is still constant.

## 3. Application to Quantum field theory

The Klein-Gordon equation corresponding to (9) is

$$\hbar^2 \frac{\partial^2}{\partial t^2}\Psi = c^2 \hbar^2 \nabla^2 \Psi + m_0^2 c^4 \Psi$$

and the Dirac equation is

$$E\Psi = -i\hbar c \begin{pmatrix} 0 & \sigma \\ \sigma & 0 \end{pmatrix} \nabla\Psi + i\begin{pmatrix} I & 0 \\ 0 & -I \end{pmatrix} m_0 c^2 \Psi$$

Especially, the interval $[0,\infty]$ of the momentum $p = m_0 V/\sqrt{1-V^2/c^2}$ is merely corresponding to the velocity $0 \leq V \leq c$ of tardyons in special relativity. Suppose a massive particle can exceed the light speed of light in vacuum and Equ.(8) is applied, the momentum interval should be $[\infty, m_0 c]$ to $c \leq V \leq \infty$. To include faster than light interval to describe the particle is helpful to reduce the divergent integral of quantum field theory. For example, in the simplified calculation of the Lamb shift [3] the energy difference $U'-U$ is

$$U' = -e\left\{1 + \frac{1}{6}\overline{(\Delta \mathbf{r})^2}\nabla^2 + \ldots\ldots\right\}\Phi(\mathbf{r})$$

$$U = -e\Phi(\mathbf{r})$$

$$U'-U = -\frac{e}{6}\overline{(\Delta \mathbf{r})^2}\nabla^2\Phi \tag{11}$$

$$\overline{(\Delta \mathbf{r})^2} = \frac{2}{\pi}\frac{e^2}{\hbar c}\left(\frac{\hbar}{m_0 c}\right)^2 \int \frac{d\omega}{\omega} \tag{12}$$

The lower limit $\omega_{min} = \dfrac{m_0 e^4}{2\hbar^3 n^2}$ ($n = 1,2,3\ldots\ldots$) equals to the energy level in the atom is acceptable. As to the upper limit, however, the author introduced the cut-off frequency $\omega_{max} = \dfrac{m_0 c^2}{\hbar}$ to avoid divergence. In fact, $\int \dfrac{d\omega}{\omega} = \int \dfrac{dk}{k} = \int \dfrac{d(\hbar k)}{\hbar k}$ ($\omega = kc$) and the momentum of the photon $\hbar k$ is also related to the momentum $p$ of the electron because of momentum conservation. Consequently, the total integral corresponding to the velocity interval $V_{min} \leq V \leq c$ ($V_{min}$ is the velocity of the electron in the atom) and $c \leq V \leq \infty$ should be

$$\int_{\hbar k min}^{\infty} \frac{d(\hbar k)}{\hbar k} + \int_{\infty}^{m_0 c} \frac{d(\hbar k)}{\hbar k} = \ln \frac{m_0 c}{\hbar k_{min}} = \ln \frac{m_0 c^2}{\hbar \omega_{min}} = \ln \frac{2n^2}{\alpha^2} \tag{13}$$

which is naturally limited and the cut-off function is unnecessary. Substituting (13) and (12) into (11),

$$U'-U = \frac{4}{3}e^2\alpha\left(\frac{\hbar}{m_0 c}\right)^2 \ln \frac{2n^2}{\alpha^2}\delta(\mathbf{r})$$

In view of $|\psi(0)|^2 = \dfrac{1}{\pi n^3 a_0^3}$, the result

$$\int (U'-U)|\psi(\mathbf{r})|^2 d^3 x = \frac{4}{3}e^2\alpha\left(\frac{\hbar}{m_0 c}\right)^2 |\psi(0)|^2 \ln \frac{2n^2}{\alpha^2} = \frac{8}{3\pi}\alpha^3 \frac{R\hbar}{n^3}\ln \frac{2n^2}{\alpha^2} \xrightarrow{n=2} 1040\,MHz$$

is approximate to the experimental data $1057\,MHz$.